# Magnetic field effects on the quantum spin liquid behaviors of $NaYbS_2$

Jiangtao Wu[1†], Jianshu Li[2,3†], Zheng Zhang[2,3†], Changle Liu[4], YongHao Gao[4], Erxi Feng[5,6], Guochu Deng[7], Qingyong Ren[1], Zhe Wang[8,9], Rui Chen[10], Jan Embs[11], Fengfeng Zhu[5], Qing Huang[12], Ziji Xiang[13], Lu Chen[13], Yan Wu[6], E.S. Choi[14], Zhe Qu[8], LuLi[13], Junfeng Wang[10], Haidong Zhou[12,14], YixiSu[5], Xiaoqun Wang[1], Gang Chen[15,4,16], Qingming Zhang[17,2*] and Jie Ma[1,10*]

**Abstract**

Spin-orbit coupling is an important ingredient to regulate the many-body physics, especially for many spin liquid candidate materials such as rare-earth magnets and Kitaev materials. The rare-earth chalcogenides $NaYbCh_2$ (Ch = O, S, Se) is a congenital frustrating system to exhibit the intrinsic landmark of spin liquid by eliminating both the site disorders between $Na^+$ and $Yb^{3+}$ ions with the big ionic size difference and the Dzyaloshinskii-Moriya interaction with the perfect triangular lattice of the $Yb^{3+}$ ions. The temperature versus magnetic-field phase diagram is established by the magnetization, specific heat, and neutron-scattering measurements. Notably, the neutron diffraction spectra and the magnetization curve might provide microscopic evidence for a series of spin configuration for in-plane fields, which include the disordered spin liquid state, 120° antiferromagnet, and one-half magnetization state. Furthermore, the ground state is suggested to be a gapless spin liquid from inelastic neutron scattering, and the magnetic field adjusts the spin orbit coupling. Therefore, the strong spin-orbit coupling in the frustrated quantum magnet substantially enriches low-energy spin physics. This rare-earth family could offer a good platform for exploring the quantum spin liquid ground state and quantum magnetic transitions.

## 1 Introduction

Quantum spin liquid (QSL) is a long-range entangled “quantum liquid” state of interacting spins [1–3], which is believed not only to be a platform for obtaining the original driving force of the high-temperature superconductivity but also to be a basic unit for the quantum data storage and computation. Although there has been much theoretical effort on different models of QSL, a firm experimental establishment of this exciting and exotic phase in a real compound is still lacking. Although limitedbehaviors have been confirmed andactivelyinvestigated on a few spin liquid candidate compounds, such as the organic materials [4–7] $\kappa$-(BEDT-TTF)$_2$Cu$_2$(CN)$_3$ and EtMe$_3$Sb[Pd(dmit)$_2$]$_2$, kagomé herbertsmithite [8, 9] ZnCu$_3$(OH)$_6$Cl$_2$, the cluster Mott insulator [10, 11] 1T-$TaS_2$ in the commensurate charge density wave phase, the honeycomb Kitaev materia [12–16], and the pyrochlore spin ice materials [17–27], there is always some unexpected factors observed in these candidate materials, such as the distorted lattice, relatively low-energy scale of the spin interactions, anti-site ionic disordering, and restricted experimental resolution, and the intrinsic magnetic behaviors might be shaded. Therefore, investigating an ideal compound and revealing the intrinsic quantum and magnetic properties related to QSLs is crucial to both material and physics research.

* Correspondence: qmzhang@iphy.ac.cn; jma3@sjtu.edu.cn
[17] School of Physical Science and Technology, Lanzhou University, Lanzhou 730000, China
[1] Key Laboratory of Artificial Structures and Quantum Control, Shenyang National Laboratory for Materials Science, School of Physics and Astronomy, Tsung-Dao Lee Institute, Shanghai Jiao Tong University, Shanghai 200240, China
Full list of author information is available at the end of the article † Equal contributors

Previously, part of us and many others have worked on a rare-earth triangular lattice magnet $YbMgGaO_4$ [28–48] where the $Yb^{3+}$ ions form a perfect triangular lattice with spin-orbit-entangled effective spin-1/2 local moments. Thus, $YbMgGaO_4$ is likely to be the first spin liquid candidate in the strong spin-orbit-coupled Mott-insulators with odd electron fillings [29, 30, 49] and goes beyond the conventional Oshikawa-Hastings-Lieb-Schultz-Mattis theorem [50–53] that requires the U(1) spin rotational symmetry. However, the potential issue of $YbMgGaO_4$ is the Mg-Ga disorder in the non-magnetic layers. To what extent the non-magnetic disorder in the non-magnetic layer would impact the intrinsic magnetic properties of the system may require further scrutiny. Instead of delving with $YbMgGaO_4$, we turn to a different route. This is motivated by the robustness of QSLs and the principle of universality. If the QSL ground state is relevant for $YbMgGaO_4$, similar QSL state could likely be realized in other triangular lattice rare-earth magnets. On the other hand, other triangular lattice rare-earth magnets could potentially eliminate the non-magnetic disorder issue that was suggested for $YbMgGaO_4$. Indeed, a series of rare-earth chalcogenides was recently identified [54, 56–67] as candidates for QSLs, and apparently, there is no structural disorder like the Mg/Ga ions in $YbMgGaO_4$. Therefore, we combine the experimental measurements and theoretical analysis and propose $NaYbS_2$ likely to be a gapless QSL candidate. The field- and temperature-dependence of magnetization, specific heat, neutron diffraction, and inelastic neutron scattering measurements demonstrate several field-induced magnetic transitions in this QSL ground state system. Moreover, we provide extra theoretical results and suggestions for further experimental verification of the QSL state in the isostructural compounds.

## 2 Samples and experimental techniques

The polycrystalline and single crystals of $NaYbS_2$ were synthesized by the high-temperature solid-state reaction and the NaCl-flux method, respectively [54]. Around 0.5 g single crystals of $NaYbS_2$ were co-aligned in the (HHL) scattering plane on a copper plate, and the single-crystal neutron diffraction data were collected at the Australia Nuclear Science and Technology Organisation (ANSTO), using a cold neutron triple-axis diffractometer, Sika [55]. An 8 T vertical magnetic field cryostat with a dilution refrigerator insert was applied to collect magnetic field data at 0 T, 3 T, 4 T, 5 T, and 8 T at 0.1 K. The magnetic field was applied along [1 – 1 0], and the data were collected with final energy as $E_f \sim 5$ meV. Rietveld refinement was performed with the SARAh software and the magnetic symmetry approach at the Bilbao Crystallographic Server. The fitted data are shown in Fig. 1 at 300 K and 1.6 K. The inelastic neutron-scattering powder data were collected at the Swiss Spallation Neutron Source (SINQ), the Paul Scherrer Institut (PSI), using a disc chopper spectrometer, FOCUS, in a 10 T magnetic field with a dilution insert. Incident neutrons with 5 Å wavelength were used with the medium-resolution chopper setting.

Specific heat measurements were performed for sintered pellets of $NaYbS_2$ down to 0.3 K at zero-field on a physical property measurement system (PPMS, Quantum Design), and the He-3 refrigerator insert was applied for temperatures below 1.8K. Moreover, data under 3T, 4T, 5T, 6T, 7 T, 8T, and 9T applied fields were also collected. The magnetic signal of specific heat was determined by subtracting the data of non-magnetic compound $NaLuS_2$ between 1.8 K and 30 K. The calculated magnetic entropy was obtained by integrating $C_{mag}/T$ between 0.3 K and 20 K.

Magnetic properties were measured by using several different instruments. Isothermal DC magnetization up to 9 T was collected on a PPMS with a vibrating sample magnetometer insert, and isothermal AC susceptibility in applied fields up to 7 T from 0.5 K to 20 K was obtained on a magnetic property measurement system (MPMS3, Quantum Design). The DC magnetic susceptibilities versus H up to 18 T at 0.8 K were collected at Wuhan National High Magnetic Field Center. Torque magnetometry was performed at the University of Michigan using a self-built capacitive cantilever setup mounted inside a dilution refrigerator.

## 3 Neutron diffraction and magnetic structure

The rare-earth ions in the chalcogenides $NaYbCh_2$ (Ch = O, S, Se) have an ideal triangular lattice structure with the magnetic $Yb^{3+}$ ions localized at the center of chalcogenide octahedra, Fig. 1(a), and the $Yb^{3+}$ triangular layers are well separated by the non-magnetic $NaCh_6$ octahedra. Since the difference between the $Na^+$ and $Yb^{3+}$ ionic sizes is large, the anti-site disorder is eliminated. The single-crystal XRD data have confirmed that the anti-site disorder between them is less than 4%. $NaYbCh_2$ has a perfect triangular lattice, and the intrinsic properties of the QSL, if there exists any, would be more clearly presented. We perform the neutron powder diffraction on $NaYbS_2$ and compare the results at 50 mK and 300 mK, Fig. 1(b). From the labeled lattice reflections, the overplotting agreement suggests no long-range magnetic ordering down to 50 mK in this system. The space groups of $NaYbCh_2$ is $R\bar{3}m$, which is same as $YbMgGaO_4$ and different to the $P$-lattice type triangular oxides with a rotation axis $C_3$ or $C_6$ of the $Z_3$ symmetry, such as $Ba_3CoSb_2O_9$ and $K_2Mn_3(VO_4)_2CO_3$ [68, 69]. Meanwhile, the external magnetic field could break the degenerated ground state and induce complicated magnetic structures ([57] and see Additional file 1).

To determine the magnetic structure of new phases under the field, single-crystal neutron diffraction was applied,

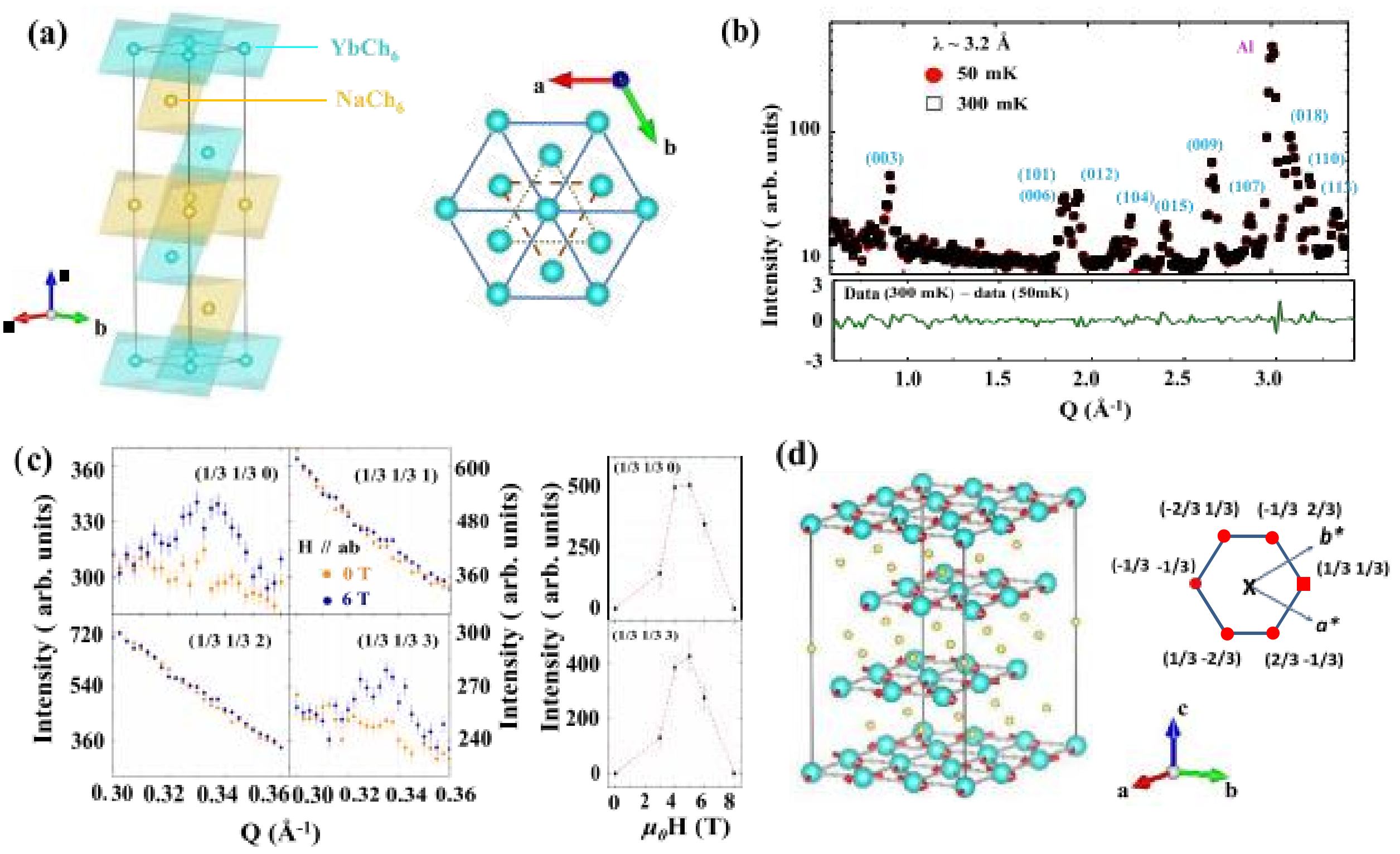

**Figure 1** Crystal structure and powder diffractions of $NaYbS_2$ . (**a**) The lattice structure and the different triangular-layers of $NaYbS_2$ . The three-layers of triangles are marked with solid lines, dot lines and broken lines, respectively. (**b**) The comparison of the neutron powder diffractions at 0.05 K and 0.3 K under 0 T with the wavelength of 3.2 Å$^{-1}$ . The lattice diffractions have been labeled and the olive line presents the difference. (**c**)The comparison of the neutron reflections at (1/3, 1/3, L) (L: 0, 1, 2, and 3) between 0 T and 6 T near 0. 1 K and the magnetic field-dependence of (1/3, 1/3, 0) and (1/3, 1/3, 3), respectively. The vertical error bars indicate the statistical errors of one standard deviation. The reflections of (1/3, 1/3, 0) and (1/3, 1/3, 3) are clearly obtained at 6 T, while (1/3, 1/3, 1) and (1/3, 1/3, 2) are too weak to be observed. (**d**) Schematics of the Brillouin zones with equivalent K-points denoted by red dots, and the magnetic structure of $NaYbS_2$ between 3 T and 6 T

Fig. 1(c). Although up-up-down (uud) plateau phase has been reported to be induced by the magnetic field in the isostructure $NaYbO_2$ [57]and $NaYbS_2$ has the similar powder neutron diffraction spectra except an extra reflection at $\boldsymbol{q} \approx 1.2429$ Å$^{-1}$ (see Additional file 1), surprisingly, the reflections of (1/3, 1/3, 1) and (1/3, 1/3, 2) were too weak to be observed under the field, which is different to the powder data of $NaYbO_2$ ([57] and see Additional file 1). Based on the program SARAh [70] and the magnetic symmetry approach at the Bilbao Crystallographic Server [71], the magnetic $Yb^{3+}$ ions spiral upwards with a 120° state and the equivalent reflections of (2/3, – 1/3, 1) and (2/3, – 1/3, –2) are instead of (1/3, 1/3, 1) and (1/3, 1/3, 2), Fig. 1(d). Therefore, the magnetic wave vector is (1/3, 1/3, 0). As the magnetic field increased, the reflections of (1/3, 1/3, 0) and (1/3, 1/3, 3) disappeared above 8 T. Since this high field phase does not have the wave vector of (1/3, 1/3) in $ab$-plane, it is disappointed to minimize the possibility of reported $uud$ plateau in $NaYbO_2$ .

Compared to the $M(H)$ measurements with the van Vleck correction, the saturated magnetization is $1.0\mu_B$ per $Yb^{3+}$ ion and the moment of the intermediate phase is $0.5\mu_B$ per $Yb^{3+}$ ion with a constant dM/dH, Fig. 2(a). The plateau phase has been predicted by Ye et al. as an up-up-up-down (uuud) state and breaks either $Z_3$ orientational symmetry or $Z_4$ sublattice symmetry with the competition of the magnetic field and quantum fluctuations [72]. Therefore, the subsequent state above the 120° state is an uuud state, which has been realized on the triangular lattice antiferromagnet $Fe_{1/3}NbS_2$ [73] and $YbMgGaO_4$ [74], and the stabilization of such a state requires further neighbor exchange interaction. A recent study on the diamond lattice antiferromagnet $LiYbO_2$ does suggest the importance of the second neighbor exchange interaction [75]. Moreover, the magnetic wave vector of $uuud$ phase shifts from (1/3, 1/3, 0) to (0, 1/2, 0) in $AgNiO_2$ [76, 77]. On further increasing the magnetic field, the $uuud$ phase undergoes a second-order phase transition to a high-field oblique phase, commonly to be observed in triangular lattices [72].

## 4 Thermodynamic data at finite temperature and phase diagram

To reveal the magnetic phases in detail, the DC and AC magnetizations and the torque magnetometry were measured. For zero field, the Curie-Weiss temperatures of $\Theta_{CW,\perp}$ = – 13.5 K and $\Theta_{CW,\parallel}$ = –4.5 K for the in-plane and out-of-plane susceptibility measurements [56]present the related frustration parameters as $f$ > $|\Theta_{CW,\perp}|$/(0.05 K) ≈ 270 for the in-plane value and $f$ > $|\Theta_{CW,\parallel}|$/(0.05 K) ≈ 90 for the out-of-plane value. Meanwhile, the larger in-plane Curie-Weiss temperature expects to enhance the spin correlation between the in-plane spin components and contribute mostly to the diffusive scattering peak. The exchange energy scale of $NaYbS_2$ is a couple times larger than the ones in $YbMgGaO_4$, consistent with the enhanced Curie-Weiss temperatures. This advantage allows a broader temperature window to explore the QSL physics in this system. Furthermore, it is noted that the magnetic susceptibility in the zero field limit is constant. Although it seems to be consistent with a spinon Fermi surface state, this is expected from the fact that the total magnetization is not a conserved quantity. Thus, the constant spin susceptibility cannot be used to identify the QSL ground state for this material and other materials. Unlike $NaYbO_2$, $NaYbS_2$ demonstrates quite rich phases with a higher saturated magnetization field (up to ~ 16.0 T), Fig. 2: there are four transition fields for DC susceptibility at 0.8 K, ~ 3.3 T, 6.1 T, 10.2 T, and 14.8 T, respectively, and are consistent with the transition fields by the AC susceptibility very well (see Additional file 1). Figure 2(b) presented the torque magnetometry measurements by applying the field in the $ab$-plane. The $dM/dH$ curves between 30 mK and 600 mK show two peaks on the lower and upper critical fields. As the temperature increases, this derivative becomes weaker and finally disappears around $T$ =1 K. Finally, the sublattice spins are aligned along the applied field and saturated.

The low-temperature specific heat measurements on $NaYbS_2$ were performed down to 0.3 K at zero-field, Fig. 2(c). As the neutron powder diffraction measurement, no signature of magnetic transition is observed, and a broad peak occurs at about 1 K, a typical phenomenon of the QSL candidate materials. The lattice phonon contribution to the specific heat is obtained from the isostructural non-magnetic material $NaLuS_2$ and could be almost negligible below 3 K, Fig. 2(c). The entropy release for $NaYbS_2$ from 0.3 K to 20 K saturates up to 92% of the $R\ln 2$entropy (where $R$ is the ideal gas constant), and an effective spin-1/2 description of the $Yb^{3+}$ local moments is obtained. Unlike $YbMgGaO_4$, $C_P/T$ of $NaYbS_2$ does not diverge at low temperatures, while the divergence in $YbMgGaO_4$ is interpreted as thesignature of U(1) gaugefluctuations [29, 30]. The intercept of $C_P/T$ on the $T$ = 0 axis is slightly larger than the one for $NaYbO_2$, and the specific heat fits well with a $T^2$ behavior below 0.7 K, the inset in Fig. 2(c), which is consistent with a gapless QSL and the Lorentz invariance

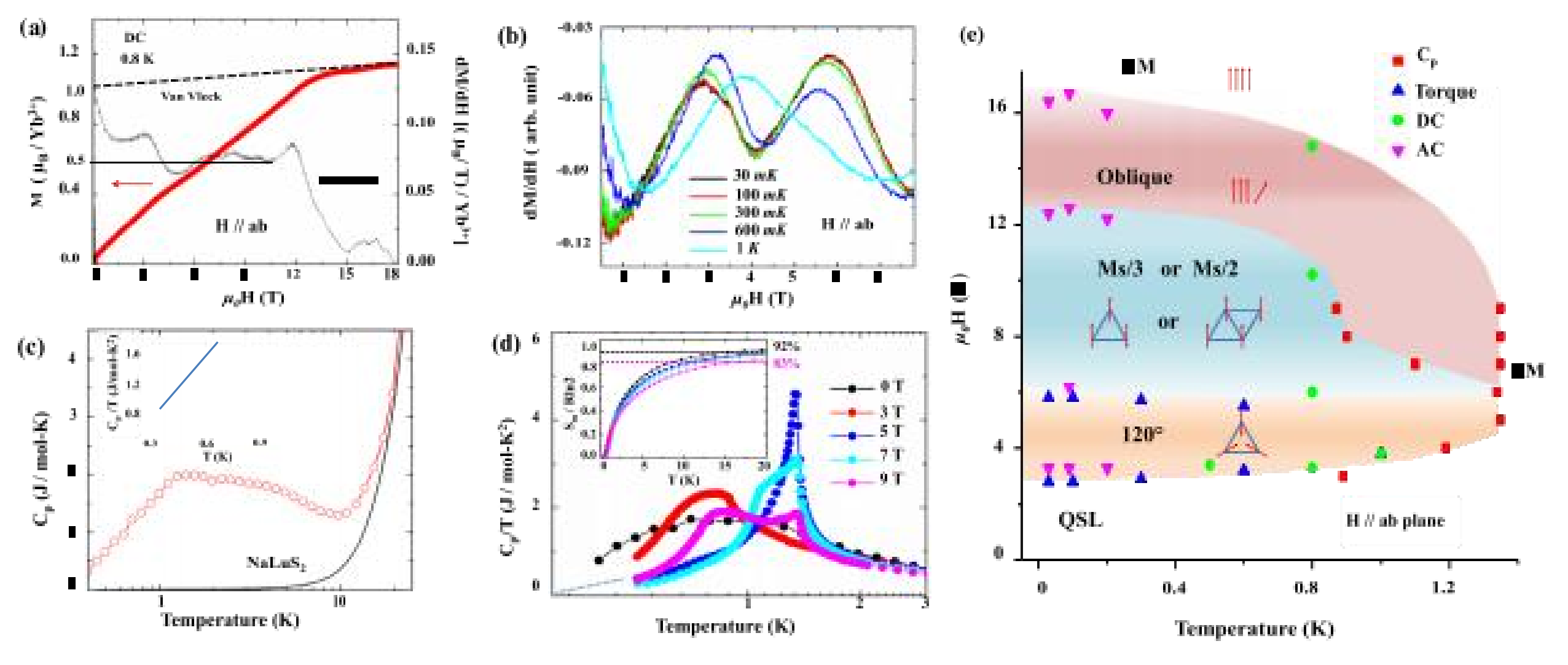

**Figure 2** The magnetization, torque magnetometry, specific heat and the phase diagram with H//ab-plane of $NaYbS_2$. (**a**) DC magnetic susceptibility versus H and the related derivative dM/dH versus H at 0.8 K. The anomaly fields are ~ 3.3 T, 6. 1 T, 10.2 T and 14.8 T. The dash lines present the saturated magnetization of 1.0$\mu_B$ per Yb ion with van Vleck correction and the intermediate Ms/2 anomaly-phase of 0.5$\mu_B$ per Yb ion, respectively. (**b**) The temperature-dependence of the torque magnetometry in DC field. The fields marked by the black arrows are suggested as the magnetic phase transition fields. (**c**) The specific heat of $NaYbS_2$ and $NaLuS_2$, respectively. Inset presents the linear relationship of $C_P$/T vs T below 0.7 K. (**d**) The magnetic field-dependence of specific heat divided by T as a function of temperature and the related magnetic entropy. (**e**) The proposed magnetic phase diagram of temperature versus magnetic field for $NaYbS_2$ with different techniques

guarantees. A residual density of states that contribute to the small intercept could be regarded [78]asthe tail of the contributions of $Yb^{3+}$ nuclear spin and becomes more and more significant below 0.1 K.

As the external magnetic field is applied, the specific heat demonstrates a series of highly tunable states, which relate to the quantum and thermal fluctuations, Fig. 2(d): a broad peak starts to be observed at 3 T, then a sharp peak appears at 5 T around ~ 1.3 K and turns to be suppressed gradually with the increasing field. Finally another broad peak is split from the sharp peak and shifts to the lower temperature at 7 T. Those transitions could also be recognized by the magnetic entropy, which decreases from 92% (0 T) to 86% (3 T), then increases 88% (5 T), and drops again to 83% (9 T). Meanwhile, the low-temperature specific heat below the peak temperature is strongly suppressed, and $C_P/T$ actually goes to zero in the zero temperature limit. Hence, the magnetic field breaks the QSL ground state and induces a range of intermediate states before the magnetic $Yb^{3+}$ ions in $NaYbS_2$ are fully saturated.

The temperature versus magnetic field diagram is plotted in Fig. 2(e). The phase transitions are determined by combining different techniques, and the various exotic states from 120° to the *uuud* Ms/2 anomaly-phase to oblique phases were decided. Especially, the thermal effect is realized in the system even at low temperatures, ~ 0.8 K, with the S-shape transition between the Ms/2- or Ms/3-anomaly phase and oblique phases.

## 5 Inelastic neutron scattering and gapless spin liquid

We perform the inelastic neutron scattering (INS) measurements on $NaYbS_2$ both at $H$ = 0 T and 6 T at 50 mK. This measurement contains the dynamical energy-momentum information about the magnetic excitations in the system. In Fig. 3(a), highly dispersive signals are revealed. It is proposed a U(1) spinon Fermi surface spin liquid in $YbMgGaO_4$ with a clear V-shape at Τ point [30]. Several key features can be lost for the powder sample of $NaYbS_2$ due to the lack of angular momentum information. However, there still exits a cone feature that can be distinguished at $|\boldsymbol{Q}| \approx 1.24$ Å$^{-1}$ from Fig. 3(a), which should correspond to the cone-like feature of the Dirac spin liquid. The inter-Dirac cone scattering and intra-Dirac cone scattering processes would present these characters at low energies. Meanwhile, for the spinon Fermi surface states, a large amount of low-energy intensity is expected in a wider momentum range, which is clearly incompatible with the experimental data. For example, $NaYbSe_2$ [79] is a quantum spin liquid candidate with spinon Fermi surface states. With magnetic field $H$ = 6 T, as depicted in Fig. 3(b) and the insets of constant-energy and -moment cuts in Fig. 3(c) and (d), the low-energy spectral weight is mostly transferred to higher energies, consistent with specific heat data that this field-induced state should be gapped due to the anisotropic spin interactions between the $Yb^{3+}$ local moments, which is different to $NaYbO_2$ with the bigger $S^{2-}$ ion on the Ch-site. Moreover, the temperature dependence of the energy- and moment-integrated intensities, Fig. 3(c) and (d), clearly presented that the V-shape magnetic signals start to be observed around 8 K.

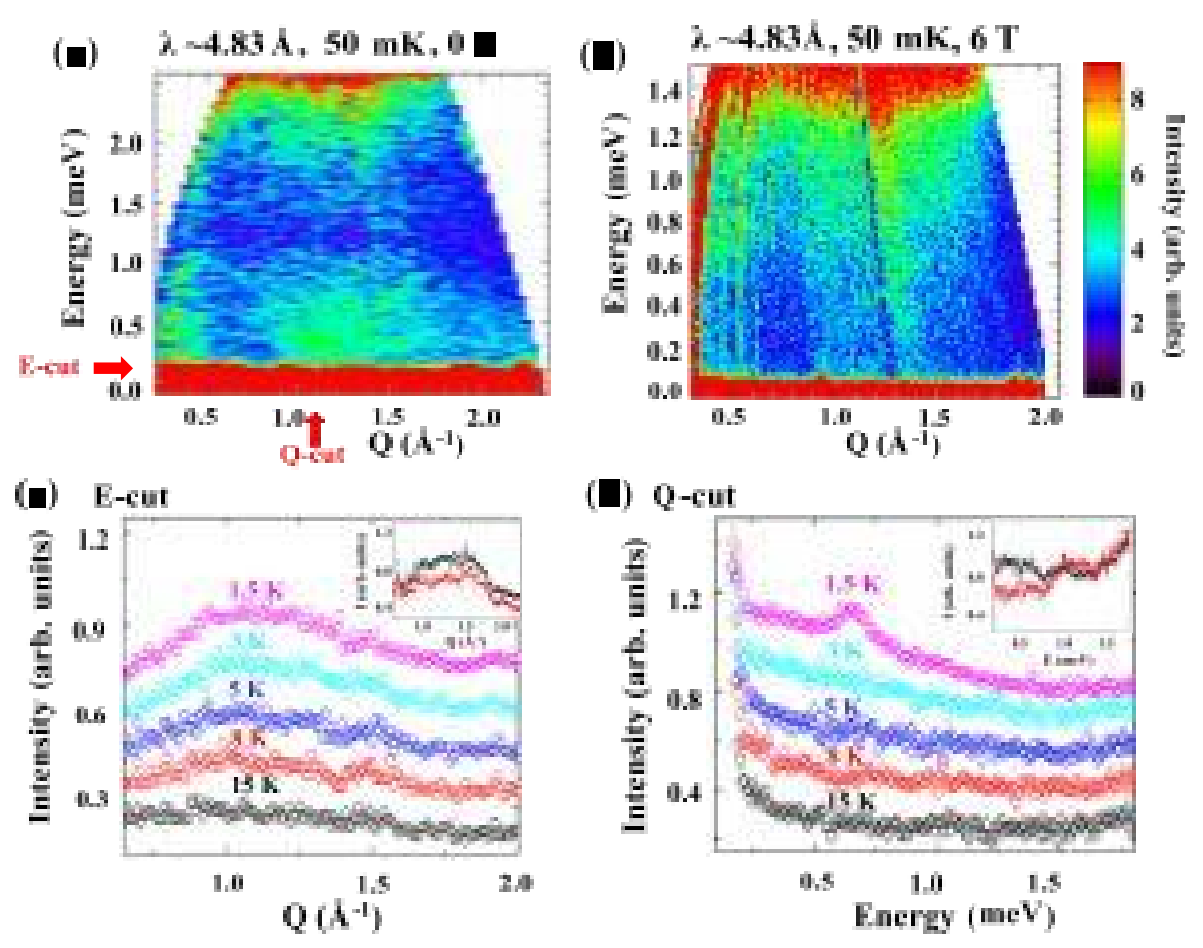

**Figure 3** Inelastic neutron scattering of $NaYbS_2$ . (**a**), (**b**) Powder-averaged inelastic neutron scattering spectra of $NaYbS_2$ at T =0.05 K, H = 0 T and 6 T, respectively. There is no signal observed below 1 meV at H = 6 T and an energy gap is suggested. (**c**), (**d**)The temperature dependence of the energy-integrated intensity (from 0.2 meV to 1.0 meV) and momentum-integrated intensity (from 0.9 Å$^{-1}$ to 1.3 Å$^{-1}$), respectively. Although there is no magnetic long-range ordering down to 50 mK at 0 T, the magnetic dynamics starts around 8 K. The insets present the related magnetic field dependence of $NaYbS_2$ at 0.05 K. The constant-energy cut (**c**) demonstrates that the magnetic signal at 6 T is still between 0.9 Å$^{-1}$ and 1.3 Å$^{-1}$;the constant-Q cut (**d**) demonstrates that there is no magnetic signal below 0.8 meV at 6 T and the gap is generated. Error bars correspond to one standard error, and the high intensity at the upper energy limits is the artefact from the background of the spectrometer and magnet

Furthermore, $NaYbS_2$ was also suggested as a gapless quantum spin liquid by longitudinal field (LF) muon spin relaxation (μSR) [80]. At 0.1 K, an indicator as the spin relaxation rate, $\lambda_{LF}$, was applied to describe the spin dynamics and obtained from LF-μSR experimental data of $NaYbS_2$ . As the magnetic field increased to 1000 G, $\lambda_{LF}$ was almost close to zero, significantly different from the LF-μSR spin relaxation rate $\lambda_{LF}$ of quantum spin liquids with spinon Fermi surface state. For example, the quantum spin liquid material $NaYbSe_2$ demonstrated a spinon Fermi surface state [79], and the spin relaxation rate $\lambda_{LF}$ maintained a constant value of ~0.2 $\mu s^{-1}$ at 0.1 K and the magnetic field of 1000 G [80]. Although both $NaYbS_2$ and $NaYbSe_2$ had magnetic ground states with quantum spin liquid states, and the crystallographic structures were close, the spinon excitation characteristics were signifi-cantly different. Therefore, the substitution of the oxychloride element effect was clearly demonstrated, and the crystal electric field played an important role in regulating quantum spin liquids [81].

## 6 Summary

Comparing to $NaYbO_2$, the bigger $S^{2-}$ ion on the Ch-site not only prevents crystallographic site-mixing, but also inceases the interlayer distances via the mediating cation, therefore the $Yb^{3+}$ layer distance along the $c$ axis is reduced significantly in $NaYbS_2$ with the slightly larger $a$ axis and leads to the $c/a$ ratio for $NaYbS_2$ (5.1) is more closer to $YbMgGaO_4$ (7.4) than $NaYbO_2$ (4.9). Rather than a pure two-dimensional model with anisotropic exchanges, a more precise theoretical analysis should naturally include other factors, such as the interlayer couplings [57]. Although the spins in the oxide are expected to be more localized than in sulfide, the quantum effects from both spin- and thermal-fluctuations in the latter are more complicated. Additionally, the ground state of $NaYbS_2$ can be driven into a magnetically ordered state in intermediate magnetic fields. Due to the strong easy-plane exchange anisotropy of $NaYbS_2$, the numerical studies suggest a canted 120° state or an incommensurate state rather than an $uud$ state in $NaYbO_2$, which is also confirmed by the magnetization and neutron diffraction measurements.

Our experiments demonstrate that the nearly ideal triangular lattice of Yb ions in strongly spin-orbit-coupled materials $NaYbCh_2$ can realize various exotic ground states. Especially, both the thermodynamic and the neutron scattering measurements suggest $NaYbS_2$ realizes a gapless spin liquid state. According to the fermion doubling theorem, there cannot be a single Dirac cone in a lattice system, and the cones must at least come in pairs. INS experiments measure the dynamical spin structural factor, which corresponds to the particle-hole pair of spinon excitations, and the intra-cone scattering will contribute to low energy spin excitations near $\Gamma$ point. In contrast, the inter-cone scattering corresponds to low energy spin excitations at finite momentum [30, 49]. Furthermore, the scenario of staggered $\pi$-flux Dirac spin liquid would double the unit cell of spinons. This results in an enhanced periodicity of the dynamical spin structure factor despite the lack of magnetic ordering, which can be identified as a sharp feature to distinguish this peculiar fractionalized state from trivial spin glass. Although the density-of-state decided the moment-dependence of the spin dynamics and no gap above 0.1 meV, the clear feature of enhanced periodicity is smeared out due to lack of angular resolution, and more experimental efforts on $NaYbS_2$ single crystals are highly desired to distinguish if this system really hosts this $\pi$-flux gapless (or Dirac) spin liquid state.

## Supplementary information

**Supplementary information** accompanies this paper at https://doi.org/10.1007/s44214-022-00011-z.

**Additional file 1.** Supplementary information (PDF 315 kB)

### Acknowledgements

J.M. and X.Q.W. acknowledge additional support from a Shanghai talent program. Q.M.Z. acknowledges the support from Users with Excellence Program of Hefei Science Center and High Magnetic Field Facility, CAS and the synergetic Extreme Condition User Facility (SECUF), CAS. Q.H. and H.Z. thank the support from NSF-DMR-2003117. F.Z., E.F. and Y.S. thank the support of the OCPC-HGF Postdoctoral Fellowship. The torque magnetometry work at Michigan was supported by the U.S. Department of Energy (DOE) under Award No. DE-SC0020184. A portion of this work was performed at the National High Magnetic Field Laboratory, which is supported by the National Science Foundation Cooperative Agreement No. DMR- 1644779 and the State of Florida.

### Funding

This work is supported by the Ministry of Science and Technology of China (Grant No. 2022YFA1402700, 2018YFGH000095), the NSF of China (Grant No. U2032213, 11774223, 12274186, 11774352, 11974244, U1832214, and U1932215), the interdisciplinary program Wuhan National High Magnetic Field Center (Grant No. WHMFC 202122), Huazhong University of Science and Technology, and the Research Grants Council of Hong Kong with General Research Fund Grant No. 17303819 and the Strategic Priority Research Program of the Chinese Academy of Sciences (Grant No. XDB33010100). Open Access funding provided by Shanghai Jiao Tong University.

### Availability of data and materials

All data generated or analyzed during this study are included in this article and its supplementary information files.

## Declarations

### Competing interests

The authors declare no competing interests.

### Author contribution

QMZ and JM conducted the study. JSL and ZZ grew NaYbS2 polycrystalline samples and single crystals. GCD, YW and EXF carried out the neutron powder diffraction of NaYbS2 and analyzed the data. ZW, ZQ, RC, JFW, QH, ESC, and HDZ performed magnetization measurements. ZX, LC and LL performed torque magnetometry measurements. JTW, QH, HDZ and JM performed specific heat measurements. JTW, QR, FFZ, EXF, JE and ES carried out inelastic neutron scattering of NaYbS2 and analyzed the data. GC performed the Monte Carlo calculation of the proposed spin model. JTW, ZZ, GC, QMZ, and JM prepared the manuscript and the supplementary materials. All authors read and approved the final manuscript.

### Author details

[1] Key Laboratory of Artificial Structures and Quantum Control, Shenyang National Laboratory for Materials Science, School of Physics and Astronomy, Tsung-Dao Lee Institute, Shanghai Jiao Tong University, Shanghai 200240, China. [2] Beijing National Laboratory for Condensed Matter Physics, Institute of Physics, Chinese Academy of Sciences, Beijing 100190, China. [3] Department of Physics, Renmin University of China, Beijing 100872, China. [4] State Key Laboratory of Surface Physics and Department of Physics, Fudan University, Shanghai 200433, China. [5] Jülich Centre for Neutron Science (JCNS) at Heinz Maier-Leibnitz Zentrum (MLZ), Forschungszentrum Jülich, Lichtenbergstrasse 1, 85747 Garching, Germany. [6] Neutron Scattering Division, Oak Ridge National Laboratory, Oak Ridge, Tennessee 37831, USA. [7] Australian Centre for Neutron Scattering, Australian Nuclear Science and Technology Organisation, New Illawarra Road, Lucas Heights, NSW 2234, Australia. [8] Anhui Key Laboratory of Condensed Matter Physics at Extreme Conditions, High Magnetic Field Laboratory, Chinese Academy of Sciences, Hefei, Anhui, 230031, China. [9] University of Science and Technology of China, Hefei, Anhui 230026, China. [10] Wuhan National High Magnetic Field Center, Huazhong University of Science

and Technology, Wuhan 430074, China. [11] Laboratory for Neutron Scattering and Imaging, Paul Scherrer Institute, CH-5232 Villigen, Switzerland. [12] Department of Physics and Astronomy, University of Tennessee, Knoxville, Tennessee 37996, USA. [13] Department of Physics, University of Michigan, Ann Arbor, 450 Church Street, Ann Arbor, Michigan 48109, USA. [14] National High Magnetic Field Laboratory, Florida State University, Tallahassee, FL 32310, USA. [15] Department of Physics and HKU-UCAS Joint Institute for Theoretical and Computational Physics at Hong Kong, The University of Hong Kong, Hong Kong, China. [16] Collaborative Innovation Center of Advanced Microstructures, Nanjing University, Nanjing, 210093, China. [17] School of Physical Science and Technology, Lanzhou University, Lanzhou 730000, China.

## Publisher's Note